\documentclass[11pt,twoside]{article}
\usepackage{./asp2014}
\usepackage{color}

\def\hi{H{\small I}}

\hypersetup{
 colorlinks= true, 
 urlcolor  = black, 
 linkcolor = red, 
 citecolor = blue 
} 

\aspSuppressVolSlug
\resetcounters

\begin{document}

\title{Radio Jet-ISM Feedback on Sub-galactic Scales}

\author{Kristina Nyland$^1$, Dipanjan Mukherjee$^2$, Mark Lacy$^1$, Isabella Prandoni$^4$, Jeremy J. Harwood$^3$, Katherine Alatalo$^5$, 
Geoffrey Bicknell$^6$,
and Bjorn Emonts$^1$
\\
\affil{$^1$National Radio Astronomy Observatory, Charlottesville, VA, 22903; \email{knyland@nrao.edu, mlacy@nrao.edu}}
\affil{$^2$Dipartimento di Fisica Generale, Universita degli Studi di Torino, Via Pietro Giuria 1, 10125 Torino, Italy; \email{dipanjan.mukherjee@unito.it}}
\affil{$^3$Centre for Astrophysics Research, School of Physics, Astronomy and Mathematics, University of Hertfordshire, College Lane, Hatfield AL10 9AB, UK; \email{jeremy.harwood@physics.org}}
\affil{$^4$INAF-Istituto di Radioastronomia, via P. Gobetti 101, 40129, Bologna, Italy; \email{prandoni@ira.inaf.it}}
\affil{$^5$Space Telescope Science Institute, 3700 San Martin Dr., Baltimore, MD 21218, USA; \email{kalatalo@stsci.edu}}
\affil{$^6$Australian National University, Research School of Astronomy \& Astrophysics, Cotter Rd, Weston, ACT 2611, Australia;
\email{geoff.bicknell@anu.edu.au}}
}

\paperauthor{Kristina Nyland}{knyland@nrao.edu}{}{National Radio Astronomy Observatory}{}{Charlottesville}{VA}{22903}{USA}
\paperauthor{Jeremy J. Harwood}{jeremy.harwood@physics.org}{}{ASTRON}{}{Dwingeloo}{}{7990 AA}{The Netherlands}
\paperauthor{Geoffrey Bicknell}{geoff.bicknell@anu.edu.au}{}{Australian National University, Research School of Astronomy \& Astrophysics}{}{Canberra}{ACT}{2611}{Australia}
\paperauthor{Your Name Here}{AuthorEmail@email.edu}{}{Institution}{}{city}{state}{zip}{country}

\begin{abstract}
Energetic feedback by active galactic nuclei (AGNs) plays an important evolutionary role in the regulation of star formation (SF) on galactic scales. However, the effects of this feedback as a function of redshift and galaxy properties such as mass, environment and cold gas content remain poorly understood.  Given its unique combination of frequency range, angular resolution, and sensitivity, the ngVLA will serve as a transformational new tool in our understanding of how radio jets affect their surroundings.  By combining broadband continuum data with measurements of the cold gas content and kinematics, the ngVLA will quantify the energetic impact of radio jets hosted by gas-rich galaxies as the jets interact with the star-forming gas reservoirs of their hosts.  
\end{abstract}

\section{Introduction}

AGN feedback may be driven by radiative winds launched by the accretion disks of powerful quasars or spurred by radio jets/lobes as they heat, expel, or shock their surroundings.  Observational evidence for both modes of feedback have been reported (e.g., \citealt{fabian+12}, and references therein). \citet{villar-martin+14} show that, on average, radio jets appear to be capable of producing more extreme gas outflows than accretion disk winds. However, the relative importance of each mode, and the dependence on redshift and other factors, such as galaxy mass and environment, remain open areas of research. 

We also lack a fundamental understanding of exactly how radio jets transfer energy to their surroundings, how much energy is transferred to the different gas phases, and under what conditions significant positive/negative feedback is produced.  We know that radio jets may deposit energy into their surroundings through a variety of mechanisms including heating, shocks, and/or turbulence \citep{fabian+12, alatalo+15, soker+16}, and may also directly couple to gas in their surroundings and physically expel it (e.g. \citealt{morganti+13}).  However, the details of these processes -- and under which conditions and environments different mechanisms dominate -- are poorly understood.  

Identifying jet-ISM feedback and constraining its impact on galaxy evolution requires observations with high sensitivity, angular resolution, and broad frequency coverage, and thus thus poses challenges to existing radio telescopes.  With its unprecedented design consisting of $\sim$214 $\times$ 18m antennas operating from 1 to 116~GHz with baselines out to several hundred km, the ngVLA will dramatically improve our understanding of how radio jets influence their surroundings.


\begin{figure*}[t!]
\centering
\includegraphics[clip=true, trim=0.25cm 7.15cm 0.15cm 7cm, width=\textwidth]{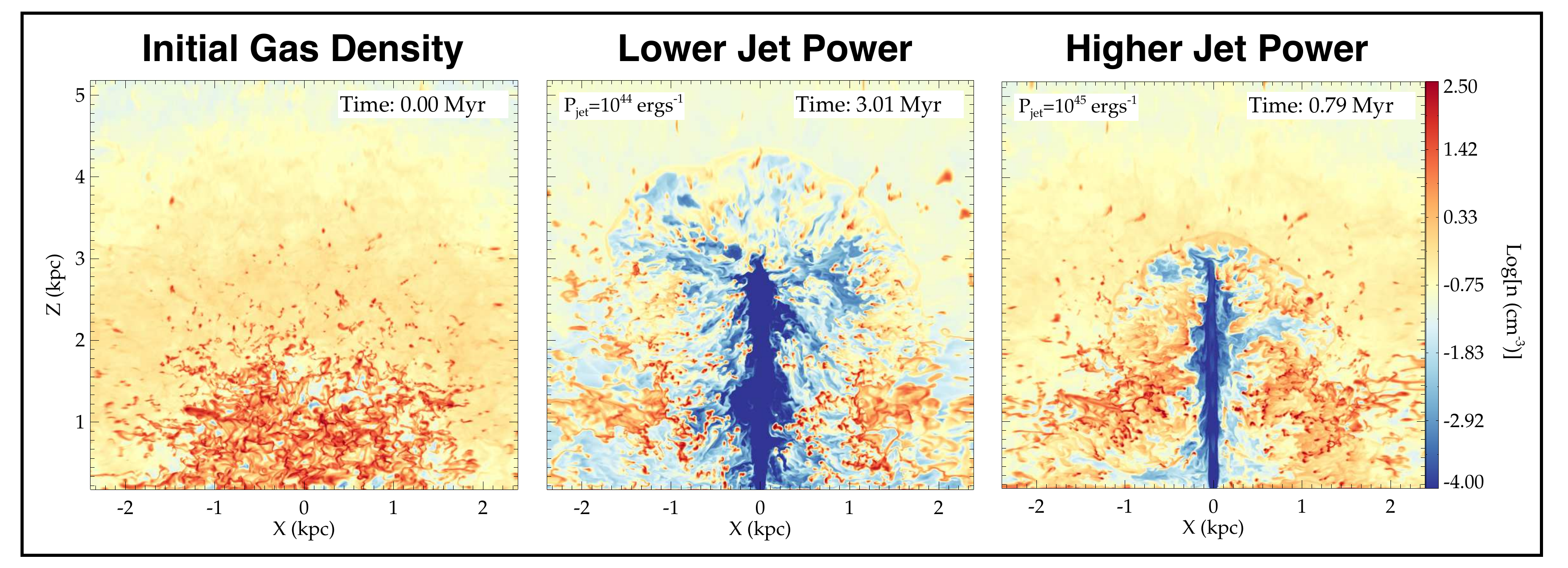}
\caption{Snapshots from the relativistic hydrodynamic radio jet simulations \citep{mukherjee+16, mukherjee+17} showing the effect on an identical initial ISM (left) made by a radio jet with $P_{\mathrm{jet}} = 10^{44}$~erg~s$^{-1}$ (center) and $P_{\mathrm{jet}} = 10^{45}$~erg~s$^{-1}$ (right). The more powerful radio jet is able to more quickly ``drill'' through the ISM of its host galaxy, while the lower-power radio jet is trapped by the ISM and able to disrupt the surrounding gas for a longer time period and over a larger volume.}
\label{fig:jet_sim}
\end{figure*}


\section{Observing Jet-ISM Feedback with the ngVLA}
The ngVLA will complement source morphologies and energetics constraints from deep, high-resolution continuum observations with spectral line data that encode information on the ISM content and conditions.  The combination of broadband continuum and spectral line imaging will allow the ngVLA to uniquely probe the energetic impact of radio jets on the ambient cold gas.  Spectral line measurements of molecular and atomic gas on comparable angular scales can be used to identify AGN-driven outflows (as well as gas inflow associated with fueling), perform detailed kinematic studies to gauge the amount of energy injected into the gas via feedback, and address the future evolutionary impact on local/global scales caused by the AGN feedback.  Molecular gas and continuum estimates of the energetics of the outflow and jet can be directly compared with state-of-the art simulations, such as those shown in Figure~\ref{fig:jet_sim}, to deeply probe the underlying feedback physics.  

Continuum + cold gas ngVLA studies would -- for the first time -- provide constraints on the prevalence and energetic importance of jet-ISM feedback in the dominant population of low-luminosity ($L_{1.4\, \rm GHz} < 10^{24}$~W~Hz$^{-1}$) AGNs residing in ``normal'' galaxies.  This is a particularly exciting prospect given recent observational evidence that lower-power radio AGNs may be able to significantly affect the interstellar medium (ISM) conditions of their hosts through feedback from sub-galactic-scale radio jets (e.g.  \citealt{alatalo+11, alatalo+15, davis+12, nyland+13, godfrey+16, querejeta+16, zschaechner+16}).  From a theoretical standpoint, recent relativistic hydrodynamic simulations of radio jets propagating through a dense ISM (\citealt{mukherjee+16, mukherjee+17}; Figure~\ref{fig:jet_sim}) provide further support for this possibility, demonstrating that while powerful radio jets rapidly ``drill'' through the ISM, lower-power jets become entrained in the ISM and are ultimately able to transfer energy over a much larger volume and for a longer period of time.  

\subsection{Sub-galactic-scale Radio Jets}
\begin{figure*}[t!]
\centering
\includegraphics[clip=true, trim=1.5cm 0.1cm 2.5cm 1cm, height=0.78\textwidth]{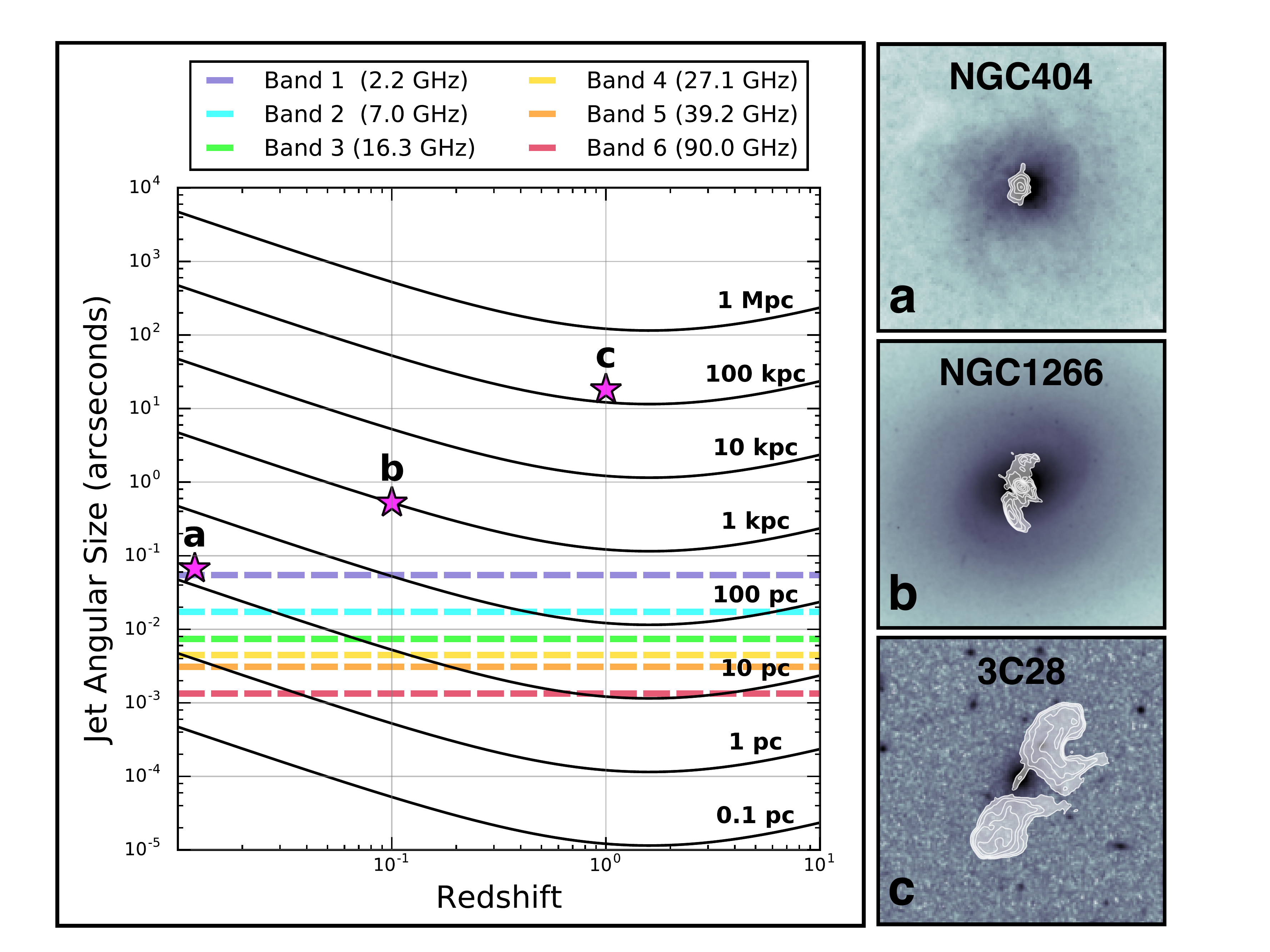}
\caption{Jet angular size as a function of redshift.  The black solid lines trace the redshift dependence of the angular extent of a jetted AGN for intrinsic jet sizes (measured from end to end along the major axis of the jet) from 0.1~pc to 1~Mpc.  The maximum angular resolution of the ngVLA at the center of each of the ngVLA bands as defined in \citet{Selina+17} is denoted by the dashed colored lines.  The magenta stars and thumbnails to the right of the main figure indicate three jetted radio AGNs representing a wide range of jet size scales:  {\bf a)} the dwarf galaxy NGC\,404 with a jet extent of 10~pc, {\bf b)} the jet-driven feedback host NGC\,1266 with a jet extent of 1~kpc, and {\bf c)} the radio galaxy 3C28 with a jet extent of 150~kpc.  The redshifts of the representative sources correspond to simulated ngVLA maps from \citet{nyland+18} at $z \approx 0$ ($D = 10$~Mpc), $z = 0.1$, and $z = 1.0$, respectively.  Figure adapted from \citet{nyland+18}.  
}
\label{fig:jet_size_z}
\end{figure*}

In Figure~\ref{fig:jet_size_z}, we illustrate the redshift dependence of the observed angular jet extent for a wide range of radio jet size scales ranging from sub-parsec jets to giant radio galaxies with Mpc-scale lobes.  The maximum angular resolution (defined as $\theta_{\mathrm{max}} = \lambda/B_{\mathrm{max}}$) for each of the proposed ngVLA observing bands (assuming $B_{\mathrm{max}} \approx 500$~km) is also shown in this figure.  Future ngVLA studies of radio jets with intrinsic extents of a few pc to a few kpc will be able to fully utilize the unique combination of angular resolution, collecting area, and frequency coverage of the ngVLA over a wide range of redshifts.  
This range of angular scales, combined with the frequency range of the ngVLA, highlights the suitability of the ngVLA for studies of jet-ISM feedback associated with lower-power radio AGNs with sub-galactic jets.  

Observations of more extended radio jets that may be engaged in feedback on intergalactic or intracluster medium scales will be possible with suitable combinations of weighting and $uv$-tapering.  The inclusion of the proposed short baseline array of $\sim$19 6m dishes in addition to the main array of 214 18m dishes would provide additional short-spacing information that would be beneficial for imaging studies of AGNs with large-scale, low-surface-brightness radio emission. 

\subsection{ISM Content and Conditions}
\label{sec:ISM_content_conditions}

\subsubsection{Ionized gas and the magneto-ionic medium}

Faraday rotation of polarized radio emission provides a unique probe of the line of sight combination of magnetic fields and ionized gas towards radio-emitting plasma. Jet-ISM interactions, particularly at high redshift, can result in material with very high Faraday rotations \citep[e.g.,][]{carilli+94}. Such high rotation measures need high frequency, high angular resolution observations to resolve the Faraday components, which can 
be mixed in (entrained) with the radio plasma, be external in a region of shocked ISM surrounding the radio source, or correspond to dense ISM within the host galaxy. These components require Faraday Synthesis \citep{brentjens+05} to disentangle them. The ngVLA, with its wide frequency coverage and high angular resolution, will be perfect for this work.

\subsubsection{Atomic Gas}
The absorption of atomic hydrogen at 21~cm against background continuum emission associated with a radio AGN provides a powerful means of directly identifying jet-driven outflows and quantifying their effect on the cold ISM (e.g., \citealt{morganti+18}).  \hi\ absorption offers a key advantage over studies of the \hi\ line in emission in terms of detectability, since the detection of \hi\ absorption is independent of redshift and depends solely on the underlying strength of the background continuum source.  In addition, the relatively low spin temperature of \hi\ of $\sim$100$-$150~K \citep{condon+16} makes the detection of emission at high angular resolution difficult or impossible due to brightness temperature sensitivity limitations (see Section~\ref{sec:caveats}).  \hi\ absorption observations, on the other hand, depend only on the brightness temperature of the background continuum source, and may therefore be performed on much smaller (e.g., milliarcsecond) scales.  
In the context of jet-driven feedback, the detection of a blue-shifted spectral component in \hi\ absorption is a signature of an outflow, which can be unambiguously distinguished from other possibilities, such as inflow or rotation. Additionally, kinematic constraints from \hi\ absorption observations probe the gas conditions by providing direct measurements of the kinetic energy of any outflow components (e.g. \citealt{nyland+13, morganti+13}) as well as characterizing the degree of turbulence \citep{lacy+17}.  

The ability of the ngVLA to observe the \hi\ line will ultimately depend on the lower frequency cutoff of its observing range.  
Assuming the ngVLA will observe down to 1.2~GHz, \hi\ studies would be limited to nearby galaxies ($0 < z < 0.1$).  \hi\ absorption surveys of bright (1.4~GHz fluxes $\gtrsim$ a few tens of mJy), nearby radio AGNs with existing radio telescopes have reported detection rates of $\sim$30\% \citep{gereb+14, maccagni+17}, suggesting that blind surveys of \hi\ absorption in even lower-luminosity systems may be possible.  The possibility of extending the ngVLA's frequency range below 1~GHz (e.g., \citealt{taylor+17}) would greatly expand the redshift range over which \hi\ would be observable with the ngVLA.  

\begin{figure*}[t!]
\centering
\includegraphics[clip=true, trim=0cm 8.1cm 0.05cm 0.09cm, width=\textwidth]{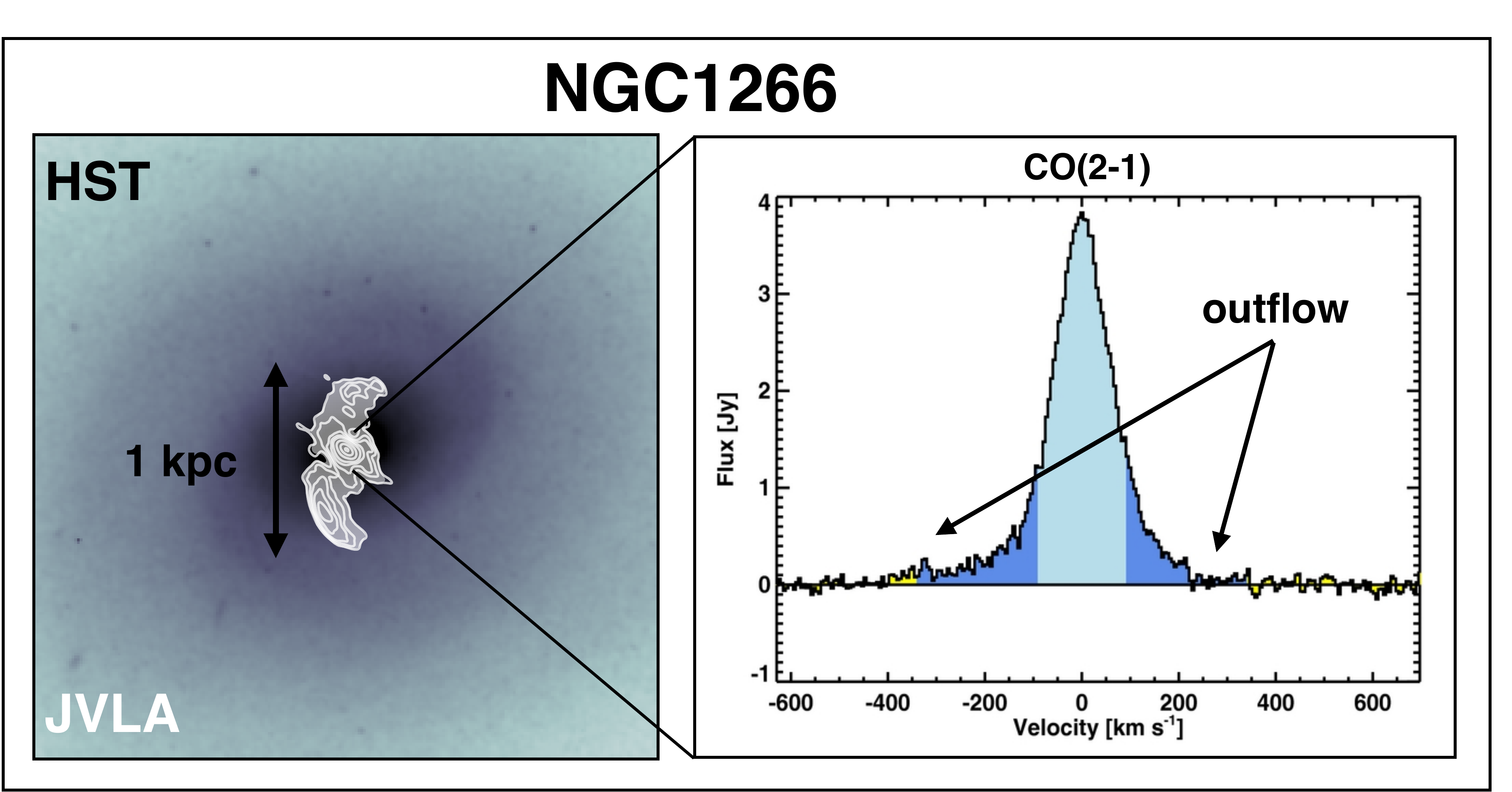}
\caption{The radio jet and outflow of the nearby galaxy NGC1266.  {\bf Left:} The background colorscale image illustrates the {\it HST} $J$-band data (WFC3, F140W; \citealt{nyland+13}) and the filled white contours trace VLA 5 GHz continuum data from \citet{nyland+16}.  {\bf Right:} The CO(2--1) data from CARMA highlights the molecular outflow that was originally identified in \citet{alatalo+11} based on the presence of excess emission in the wings of the spectrum (dark blue).}
\label{fig:NGC1266}
\end{figure*}

\subsubsection{Molecular Gas}
Identifying jet-driven molecular outflows is crucial for understanding the multiphase nature of AGN feedback (e.g., \citealt{rupke+13, emonts+14, sakamoto+14, alatalo+15,morganti+15b}).  Molecular outflows may be identified on the basis of their spectral line shapes, such as the presence of broad wings or a shifted component 
(e.g., NGC\,1266; Figure~\ref{fig:NGC1266}) 
or a P Cygni profile \citep{sakamoto+09}.  A survey of the cold gas properties of a large statistical sample of AGNs spanning a wide range of environments, host galaxy morphologies, SMBH masses, and nuclear activity classifications (e.g., high/low Eddington ratios, jetted/radio quiet, Compton thick/unobscured, etc.) would ultimately help establish an observationally-motivated model for the cosmic importance of outflows launched by active nuclei.  
We emphasize that improving our understanding of AGN feedback through ngVLA molecular gas observations will not be limited to objects with detectable outflows. 
Observations of more subtle feedback effects, such as a significant increase in the turbulence of the gas, or a substantial change in star formation efficiency/depletion time in the vicinity of the AGN,  
will also be possible (e.g. \citealt{alatalo+15,oosterloo+17}).  

The lowest energy transitions of the CO molecule, CO(1--0), CO(2--1), and CO(3--2) at rest frequencies of 115, 230, and 345~GHz, respectively, trace the total molecular gas reservoir at relatively low densities ($n_{\mathrm{H}_2} \sim 10^3$~cm$^{-3}$). 
The CO(1--0) line will be accessible to the ngVLA over the redshift ranges $0< z\lesssim 0.5$ and $z\gtrsim 1.5$.  The gap from $z \approx 0.5$ to $1.5$ is due to the high telluric opacity of molecular oxygen that precludes ground-based observations from 52 to 68 GHz.  Observations of the CO(2--1) line will be possible from $1 \lesssim z\lesssim 2$ and $z\gtrsim 3.5$, and the CO(3--2) line will be accessible over the range $2 \lesssim z\lesssim 4$ and also at $z \gtrsim 6$ (though see Section~\ref{sec:caveats} regarding important caveats).  We note that none of the low-$J$ CO lines will be observable from $z = 0.5 - 1.0$, though transitions of other species probing denser gas, such as SiO and CS, will be accessible.  For a graphical description of the redshift dependence of a wide variety of molecular gas species and transitions observable with the ngVLA, we refer readers to Figures 2 and 9 in \citet{casey+15}.

In addition, imaging of the cold molecular gas in the vicinity of the AGN can provide important clues on the fueling of the central BH. Depending on the geometry and kinematics of this gas (disks, patches or filaments),
the fueling mechanism and efficiency can be determined, in comparison with models (e.g., \citealt{gaspari+15}). 

\subsubsection{Important Caveats}
\label{sec:caveats}
The minimum excitation temperature of the CO(1--0) line is 5.53~K, and radio telescopes must therefore have sufficient brightness temperature sensitivity to detect this line. With maximum ngVLA baselines of $\sim$500~km in the north-south direction, the naturally-weighted angular resolution 
at 93~GHz would be $\sim$10~mas, corresponding to a brightness temperature sensitivity\footnote{The brightness temperature sensitivity of a point source is $\sigma_{T_\mathrm{B}} = \left( \frac{S}{\Omega_{\mathrm{A}}} \right) \frac{\lambda^2}{2k}$, where $S$ is the flux density in units of W~m$^{-2}$~Hz$^{-1}$, $k$ is the Boltzmann constant ($1.38 \times 10^{-23}$~Jy~K$^{-1}$), and $\lambda$ is the observing wavelength in meters.  The quantity $\Omega_{\mathrm{A}}$ is the beam solid angle, $\Omega_{\mathrm{A}} = \frac{\pi \theta^2_{\mathrm{FWHM}}}{4 \ln(2)}$, where $\theta_{\mathrm{FWHM}}$ is the angular resolution in radians.} of $\sigma_{T_\mathrm{B}} \sim 350$~K for an integration time of 1~hr, a channel width of 10~km~s$^{-1}$, and robust weighting \citep{Selina+17}.  
Thus, significant tapering of the data, as well as the application of new weighting schemes, will be necessary for studying jet-ISM feedback through ngVLA observations of the low-$J$ CO transitions.  

At high redshifts, the increasing influence of the cosmic microwave background (CMB) may also hinder the detectability of CO. 
The increasing CMB temperature at high redshifts\footnote{The redshift dependence of the CMB temperature follows the relation $T_{\mathrm{CMB}} = T^{z=0}_{\mathrm{CMB}} \times (1 + z)$, where $T^{z=0}_{\mathrm{CMB}} = 2.73$~K is the CMB temperature at $z = 0$.} both reduces the contrast of the CO emission against the background (particularly in cold molecular clouds with $T_{\mathrm{kinetic}} \sim 20$~K) and changes the shape of the CO spectral line energy distribution by exciting a greater proportion of higher-$J$ rotational levels  (e.g. \citealt{da_cunha+13, zhang+16}).  This may be problematic for ngVLA observations of cold molecular clouds in non-starbursting galaxies at $z \sim 4$, where $T_{\mathrm{CMB}} = 13.65$~K.  However, CMB heating is expected to be less problematic in molecular clouds in the vicinity of AGNs and jets, since AGN heating may increase the gas kinetic temperature to hundreds of degrees \citep{matsushita+98, krips+08, viti+14, glenn+15,richings+18}.  

\section{Multiwavelength Synergy}
The unique 
capabilities of the ngVLA will facilitate exciting advancements in our understanding of AGN feedback and its broader connection to galaxy evolution, particularly when combined with multiwavelength data from other state-of-the-art instruments.  In terms of synergy with current radio telescopes, observations with the Atacama Large Millimeter and Sub-millimeter Array (ALMA) at frequencies above the ngVLA's limit of 116~GHz will provide key insights into the energetic and chemical impact of jet-driven feedback on the dense gas phase of the ISM.  In the low radio frequency regime,  the Square Kilometre Array (SKA) and its pathfinders will probe the 21cm line out to higher redshifts (though at lower spatial resolution) than the ngVLA \citep{morganti+15a}.  The combination of constraints on both the atomic and molecular gas conditions from ngVLA and SKA observations will be important for studying the full impact of energetic jet-driven feedback on the ISM.  

\bibliography{ngVLA_jet_ISM_feedback_v1}  

\end{document}